# Fractional Quantum Hall Phase Transitions and Four-flux Composite Fermions in Graphene

Benjamin E. Feldman[1], Andrei J. Levin[1], Benjamin Krauss[2], Dmitry Abanin[1,3], Bertrand. I. Halperin[1], Jurgen H. Smet[2] and Amir Yacoby[1*]

[1]*Department of Physics, Harvard University, Cambridge, Massachusetts 02138, USA*
[2]*Max-Planck-Institut für Festkörperforschung, Heisenbergstrasse 1, D-70569 Stuttgart, Germany*
[3]*Perimeter Institute for Theoretical Physics, Waterloo, Ontario N2L 6B9, Canada*

*yacoby@physics.harvard.edu

**Graphene and its multilayers have attracted considerable interest owing to the fourfold spin and valley degeneracy of their charge carriers, which enables the formation of a rich variety of broken-symmetry states and raises the prospect of controlled phase transitions among them. In especially clean samples, electron-electron interactions were recently shown to produce surprising patterns of symmetry breaking and phase transitions in the integer quantum Hall regime[1-5]. Although a series of robust fractional quantum Hall states was also recently observed in graphene[6-11], their rich phase diagram and tunability have yet to be fully explored. Here we report local electronic compressibility measurements of ultraclean suspended graphene that reveal a multitude of fractional quantum Hall states surrounding filling factors $\nu$ = -1/2 and -1/4. In several of these states, we observe phase transitions that indicate abrupt changes in the underlying order and are marked by a narrow region of negative compressibility that cuts across the incompressible peak. Remarkably, as filling factor approaches $\nu$ = -1/2, we observe additional oscillations in compressibility that appear to be related to the phase transitions and persist to within 2.5% of $\nu$ = -1/2. We use a simple model based on crossing Landau levels of composite particles[12] with different internal degrees of freedom to explain many qualitative features of the experimental data. Our results add to the diverse array of correlated states observed in graphene and demonstrate substantial control over their order parameters, showing that graphene serves as an excellent platform to study correlated electron phases of matter.**

When a two-dimensional electron gas (2DEG) is subject to a perpendicular magnetic field $B$, the electronic spectrum forms a sequence of flat energy bands called Landau levels (LLs), which can accommodate one electron per magnetic flux quantum for each internal electronic state. Generally, this gives rise to incompressible quantized Hall states at integer values of the filling factor $\nu = nh/eB$, where $n$ is the carrier density, $h$ is Planck's constant and $e$ is the electron charge. In exceptionally clean samples at high magnetic field, Coulomb interactions become important and produce additional quantized Hall states at certain fractional filling factors[12-15]. These fractional quantized Hall (FQH) states can be understood in terms of so-called composite fermions (CFs), which may be described as an electron bound to an even number $m$ of magnetic flux quanta. CFs with $m = 2$ ($^2$CFs) experience a reduced effective magnetic field proportional to ($\nu$ - 1/2), and FQH states at $\nu = p/(2p \pm 1)$ are understood to arise when an integer number $p$ of $^2$CF LLs are occupied. In CF theory[12], FQH states of electrons are therefore interpreted as the integer quantized Hall effect of these new composite particles.

Like electrons, CFs can have internal quantum numbers such as spin or valley index (isospin). When more than one CF LL is occupied, ground states with different polarizations of these degrees of freedom are possible at a given filling factor, and transitions between different phases may occur when system parameters are varied. Phase transitions between FQH states with differing spin polarization have been observed in GaAs by tuning the magnitude of the magnetic field[16-19], its direction[20-25], or the applied pressure[26]. In AlAs 2DEGs, strain has been used to induce phase transitions between valley-polarized and unpolarized states[27-29].

In graphene, the electronic Hamiltonian has an approximate SU(4) symmetry arising from the spin and valley degrees of freedom. This symmetry is weakly broken due to the Zeeman effect and electron-electron scattering between valleys, which may be enhanced by (or compete with) effects of the dominant Coulomb interactions. Theoretical proposals suggest that the strengths of FQH states can be tuned in monolayer and bilayer graphene, and that transitions between different ordered phases are possible[30-32], but none have previously been observed. Here we report local electronic compressibility measurements of suspended graphene, performed using a scanning single-electron transistor (SET)[33, 34]. The ability to perform local measurements allows us to study exceptionally clean regions of graphene and reveals a succession of FQH phase transitions as well as four-flux CF states. A schematic of the measurement setup is shown in Fig. 1a (see Methods). Modulating the carrier density with a back gate and monitoring the resulting change in SET current allows us to measure both the local chemical potential μ and the local inverse compressibility of the graphene flake with a spatial resolution of approximately 100 nm. The inverse compressibility $\kappa^{-1}$ is properly defined as $n^2 d\mu/dn$, but hereafter we drop the prefactor and use the term to mean $d\mu/dn$. All the data presented below were taken at one location; however, similar behavior was observed at multiple positions along the sample (see Supplementary Information).

Figure 1b shows the inverse compressibility as a function of filling factor and magnetic field. FQH states appear as vertical incompressible peaks at $\nu$ = -1/3, -2/3, -2/5, -3/5, -3/7, -4/7, -4/9, -5/9 and -5/11, consistent with the standard CF sequence observed for $|\nu| < 1$ in previous measurements[11] (we focus on the behavior for $\nu < 0$ because the QH features are slightly better developed than for $\nu > 0$) . Surprisingly, every FQH state except $\nu$ = -1/3 exhibits a narrow magnetic field range over which the incompressible behavior is strongly suppressed and the energy gap decreases. The critical field at which this occurs increases with filling fraction denominator, and the suppression is associated with regions of sharply negative compressibility that cross each FQH state, often generating two coexisting incompressible peaks at slightly different filling factors over a small range in magnetic field (Fig. 1c). Interestingly, the negative compressibility, which indicates a decrease in the electron chemical potential as electrons are added, has similar amplitude (but opposite sign) to the incompressible peaks of the actual FQH states.

The interruptions in each incompressible peak suggest phase transitions in which the spin and/or valley polarization of electrons changes abruptly. The behavior is similar to that observed in GaAs, where transport measurements showed FQH states splitting into doublets near phase transitions[17, 18, 22, 25]. However, no dramatic features of negative compressibility were present in GaAs[19], and the inverse compressibility did not display a strong asymmetry between filling factors just above and below the FQH states[19, 35].

Several less prominent modulations in compressibility that occur close to $\nu$ = -1/2 are also visible in Fig. 1b as alternating light and dark blue features. We emphasize that they are not caused by localized states, which occur near the strongest FQH states such as $\nu$ = -2/3, but not around high-denominator states, such as $\nu$ = -4/7 (see Supplementary Information). Further insight can be gained by plotting the inverse compressibility as a function of $p$ rather than $\nu$ (Figs. 1d and e). Doing so more clearly illustrates the behavior near $\nu$ = -1/2 and reveals oscillations in inverse compressibility that persist to values of $p$ as large as 20 and magnetic fields as low as a few Tesla. We note that this behavior cannot be explained by Shubnikov de Haas oscillations of CFs, because variations in compressibility occur even at constant filling factor. The oscillations become stronger and more vertical as the magnetic field is increased, suggesting that they are associated with developing FQH states. Moreover, they seem to extend from the negative compressibility features of the phase transitions, suggesting that they result from changes in spin and/or valley polarization as magnetic field and filling factor are varied.

Signatures of phase transitions have previously been observed in compressibility measurements only at $\nu$ = 2/3 in GaAs[19], although optical and transport studies of GaAs and AlAs have revealed evidence of changes in spin or isospin polarization for filling fractions with denominators up to thirteen[20, 36]. We observe clear phase transitions up to $\nu$ = -5/9 and -5/11, with further compressibility oscillations apparent much closer to $\nu = -1/2$. Similar oscillations have not been reported in GaAs; their existence in graphene suggests a rich array of ordered electronic states and hints at a delicate energetic competition between them. Graphene therefore offers an excellent platform to study electronic interactions and their dependence on the underlying symmetry.

To gain further insight into the phase transitions, we introduce a simple model to describe CFs with internal degrees of freedom[37-40] (see Supplementary Information for details). Due to graphene's peculiar band structure, the lowest LL is already half-full at $\nu = 0$, and experiments suggest that the $\nu = 0$ state has no net spin polarization[1]. For $0 > \nu > -1$, we assume that the ground state is obtained by putting holes in the $\nu = 0$ state, which we convert to CFs by attaching two flux quanta to each hole. The holes, and hence the CFs, have two possible spin states (±) and we consider many-body states where there may be different particle densities for the two spins. Single-particle energies of the two spin states will be split by an amount proportional to $B$ due to the Zeeman effect, which favors a state where all the spins are aligned with the field. However, the SU(4) invariant part of the Coulomb interaction will typically favor states with more equal occupation[38]. Because the Coulomb interaction energies scale as $B^{1/2}$, then for fixed $\nu$, varying the magnetic field will change the relative importance of the two terms, which suggests that the experimentally observed phase transitions may be caused by changes in spin polarization, as in GaAs.

Our model applies most directly to the situation where all electrons in the ground state of $\nu = 0$ have the same valley configuration, as in the Kekule or the charge-density-wave states[41]. The antiferromagnetic state is more complicated because the constituent electron states differ in valley index as well as spin, but we expect that results for this case should be at least qualitatively similar to the case we consider. Future studies in which a tilted magnetic field is applied to the sample may help determine the spin and valley ordering of the FQH states.

Within our CF description, effects of the SU(4) invariant Coulomb interaction are modeled by a sum of the "kinetic energies" of the occupied states in the occupied CF LLs, which scale as $B^{1/2}$ for fixed orbital index $N^*$. A schematic diagram of CF LL energies $E^{\pm}_{\mathrm{N}^*}$ and their scaling with magnetic field is shown in Fig. 2a. At certain critical magnetic fields, CF LLs with different spin and orbital degrees of freedom cross, leading to phase transitions. In our simulation, we have also broadened the CF LLs by a fixed amount of disorder $\delta n$. We calculate the occupation of each CF LL, and this ultimately yields the inverse compressibility as a function of density and magnetic field. The resulting simulations, which assume either a small amount of disorder or more realistic density fluctuations based on the experimentally observed widths of the FQH peaks, are shown in Figs. 2b and c, respectively.

The simulations in Fig. 2 share many characteristics with the data presented in Fig. 1. Most striking are the breaks in the incompressible peaks of FQH states with $p > 2$, and the reasonably good agreement between the critical fields of these transitions in the simulation and their experimental counterparts. In addition, the simulations clearly show regions of negative compressibility that cross from one side of the FQH state to the other as the phase transition occurs. This is qualitatively similar to the behavior that we observe, although the experimental features are much narrower. Finally, we note that the oscillations in inverse compressibility become less robust and start to curve at low magnetic fields and high $p$, similar to the experimental data. The values used for parameters in the simulation agree well with expectations based on other experimental metrics. By matching the simulation to the experimental critical fields and assuming Zeeman splitting with a $g$-factor of 2, we extract an effective mass $m^* = 0.18m_e(\nu B[\mathrm{T}])^{1/2}$, the same order of magnitude as for CFs in GaAs[16]. In addition, the density fluctuations $\delta n = 1.5\mathrm{x}10^8$ cm$^{-2}$ assumed in Fig. 2c are comparable to the widths of the FQH states we observe. Given the simplicity of the model, the agreement with experiment is remarkable, suggesting that it provides a basic framework to understand the underlying physics.

We observe only one phase transition in our data at each filling factor, which we believe corresponds to the highest-field one. This likely reflects the level of disorder in our sample, and is consistent with our theoretical model (Fig. 2c). The model predicts that further improvements in cleanliness would allow us to observe multiple phase transitions in a given filling factor over the available experimental range (Fig. 2b). We note that there are already some low-field oscillations in compressibility (*e.g.* in Fig. 1e, starting from $B = 7$ T, just to the right of $p = 5$) that are reminiscent of the diamond-shaped compressibility modulations between incompressible peaks in Fig. 2; this behavior may be a precursor to the more delicate low-field features.

Integrating the inverse compressibility with respect to carrier density gives the chemical potential and allows us to extract the steps in chemical potential $\Delta\mu_\nu$ of each FQH state; multiplying $\Delta\mu_\nu$ by the quasiparticle charge then yields the corresponding energy gaps $\Delta_\nu$. Figures 3a and 3b show measurements of inverse compressibility and chemical potential, respectively, as a function of filling factor at 11.9 T. In Figures 3c and d, we plot $\Delta\mu_\nu$ as a function of magnetic field. The energy gap is non-monotonic for several FQH states (Fig. 3d), providing further evidence of the phase transitions discussed above. This behavior becomes increasingly pronounced and the field range over which incompressible behavior is suppressed widens as filling factor denominator increases. A similar pattern occurs in the simulations of Fig. 2, and it likely results from the increasing effects of density fluctuations on CF LL width as *p* increases.

The step in chemical potential at $\nu = -1/3$ scales linearly with $B$ over the entire field range that we study. This behavior is consistent with prior studies[11], and at the highest available experimental magnetic field, the value of $\Delta\mu_{-1/3}$ reaches 14 meV, the largest reported value. However, its linear scaling is surprising because interaction-driven states typically scale as $B^{1/2}$. The behavior also contradicts the $B^{1/2}$ scaling expected from our model, although we note that the model does not include interactions among CFs, LL mixing, finite temperature effects, or the possibility of other excitations such as Skyrmions. Linear scaling with magnetic field at $\nu = 1/3$ has been theoretically predicted to arise from spin-flip excitations over an intermediate field range[42]. The energy gaps of other FQH states exhibit complex dependence on magnetic field due to the phase transitions that they undergo, but they are qualitatively similar to the behavior observed in GaAs near phase transitions[24].

In addition to the phase transitions discussed above, the exceptional quality of the measured sample reveals several FQH states belonging to the four-flux CF ($^4$CF) sequence $\nu = p/(4p \pm 1)$ and its analogue around $\nu = -1$ for the first time in graphene. Figs. 4a-c show incompressible behavior at $\nu = -1/5$, $-2/7$, $-2/9$, $-3/11$, $-5/7$ and $-6/5$. An additional weak incompressible peak occurs between $\nu = -9/7$ and $-14/9$, but the experimental uncertainty in filling factor prevents a more precise assignment (see Supplementary Information). No other $^4$CF states are visible; although the data are not shown in Fig. 4, FQH states at $\nu = -4/5$, $-9/5$ and $-12/7$ are conspicuously absent, despite the robust appearance of their counterparts near $\nu = 0$. This may reflect interesting patterns of symmetry-breaking in the lowest LL, as was observed for the $^2$CF series, but could also be caused by differing degrees of disorder at different filling factors, or by competition with other quantum Hall states, particularly near $\nu = -2$.

The extracted steps in chemical potential for several $^4$CF FQH states are plotted as a function of magnetic field in Fig. 4d. The fluctuations caused by localized states near $\nu = -1/5$ and -2/9 prevent an accurate determination of $\Delta\mu_\nu$ for these states, but all other states except for $\nu = -5/7$ scale approximately linearly (perhaps even slightly superlinearly) with magnetic field. Further study is required to determine whether the non-monotonic behavior of $\Delta\mu_{-5/7}$ reflects a phase transition at higher field or whether the state is simply competing with $\nu = -2/3$. Regardless, the appearance of $^4$CF states and phase transitions represents an important advance in sample quality that may lead to further surprises as the delicate many-body states arising from interacting Dirac fermions continue to be explored.

**Methods**

Graphene flakes were mechanically exfoliated onto a doped Si wafer capped with 300 nm of $SiO_2$. Suitable flakes were identified by optical microscopy and were electrically contacted using electron beam lithography followed by thermal evaporation of Cr/Au (3/100 nm) contacts and liftoff in warm acetone. The sample was placed in 5:1 buffered oxide etch for 90 s and dried using a critical point dryer. It was then transferred to a $^3$He cryostat and cleaned by current annealing. All measurements were performed at approximately 450 mK. The back gate voltage was limited to ±10 V to avoid structural damage to the suspended device. The sample whose data appears in this paper is a monolayer-bilayer hybrid. Its size is 3.5 μm in width (the monolayer portion is 2 μm), and 1 μm in length (*i.e.*, distance between contacts). All the local measurements reported in the main text were conducted on the monolayer side of the flake, approximately 300 nm from the monolayer-bilayer interface and 500 nm from the electrical contacts.

To fabricate the scanning SET tip, a fiber puller was used to make a conical quartz tip. Al leads (16 nm) were evaporated onto either side of the quartz rod, and following an oxidation step, 7 nm of additional Al was evaporated onto the tip to create the island of the SET. The diameter of the SET is approximately 100 nm, and it was held 50-100 nm above the graphene flake during measurements. Compressibility measurements were performed using AC and DC techniques similar to those described in refs. [33, 34]. The SET serves as a sensitive measure of the change in electrostatic potential $\delta\Phi$, which is related to the chemical potential of the graphene flake by $\delta\mu = -e\delta\Phi$ when the system is in equilibrium. In the AC scheme used to measure $d\mu/dn$, an AC voltage is applied to the back gate to weakly modulate the carrier density of the flake, and the corresponding changes in SET current are converted to chemical potential by normalizing the signal with that of a small AC bias applied directly to the sample. For DC measurements, a feedback system was used to maintain the SET current at a fixed value by changing the sample bias. The corresponding change in sample voltage provides a direct measure of $\mu(n)$.

**Acknowledgments**

We would like to thank M. T. Allen for helping to current anneal the device. We also acknowledge useful discussions with M. Kharitonov, J. K. Jain, L. S. Levitov and S. H. Simon. This work is supported by the US Department of Energy, Office of Basic Energy Sciences, Division of Materials Sciences and Engineering under Award #DE-SC0001819. JHS and BK acknowledge financial support from the DFG graphene priority programme. BK acknowledges financial support from the Bayer Science and Education Foundation. This work was performed in part at the Center for Nanoscale Systems (CNS), a member of the National Nanotechnology Infrastructure Network, which is supported by the National Science Foundation under NSF award no. ECS-0335765. CNS is part of Harvard University.

### Competing Financial Interests

We have no competing financial interests.

### Author Contributions

B. E. F., A. J. L., B. K., J. H. S. and A. Y. conceived of and designed the experiments. B. E. F. and B. K. fabricated the sample. B. E. F. and A. J. L. performed the measurements. All authors analyzed the data and wrote the paper.

## Figure Legends

**Figure 1 | Measurement setup and phase transitions. a,** Schematic of the measurement setup. The single-electron transistor (SET) is approximately 100 nm in size and is held 50-150 nm above the graphene flake. **b,** Inverse compressibility $d\mu/dn$ as a function of filling factor $\nu$ and magnetic field $B$. Phase transitions occur for all fractional quantum Hall (FQH) states except $\nu = -1/3$. **c,** Finer measurement of the $\nu = -4/7$ phase transition clearly showing the negative compressibility and peak splitting associated with the phase transition. Panels (b) and (c) have identical color scales. **d and e,** $d\mu/dn$ as a function of magnetic field $B$ and composite fermion Landau level (CF LL) index $p$, from $\nu = -p/(2p \pm 1)$. Panels (d) and (e) have identical color scales, and both show oscillations in compressibility that persist very close to $\nu = -1/2$. Principal FQH states are marked by black lines and are labeled in panels (b), (d) and (e).

**Figure 2 | Theoretical model and numerical simulation. a,** Schematic of CF LL energies $E_{\mathrm{p}}^{\pm}$ divided by and plotted against $B^{1/2}$. Crossings (black circle) between spin-up and down CF LLs (colored arrows) correspond to phase transitions in which the polarization of a given FQH state abruptly changes. **b and c,** Numerical simulations of dμ/d*n* as a function of *B* and *p*. Black ovals correspond to the black circle in panel (a). The simulations assume either minimal charge inhomogeneity (b), or more realistic density fluctuations (c). Both panels use the same color scale.

**Figure 3 | Steps in chemical potential. a,** dμ/d*n* as a function of *ν* at *B* = 11.9 T. **b,** Chemical potential relative to its value at *ν* = -1/2 as a function of *ν* at 11.9 T. The step in chemical potential $\Delta\mu_\nu$ of each incompressible state is given by the difference in chemical potential between the local maximum and minimum (green labels). **c and d,** $\Delta\mu_\nu$ of FQH states as a function of *B* at measured multiples of *ν* = 1/3 and 1/5 (c), and *ν* = 1/7 and 1/9 (d). The non-monotonic dependence on magnetic field visible for several FQH states is a consequence of the phase transitions that they undergo.

**Figure 4 | Four-flux composite fermion states. a-c,** dμ/d*n* as a function of *ν* and *B*. FQH states belonging to the four-flux CF sequence are visible as vertical incompressible peaks at *ν* = -1/5, -2/7, -2/9, -3/11, -5/7 and -6/5. **d,** Steps in chemical potential $\Delta\mu_\nu$ of the four-flux CF states as a function of *B*.

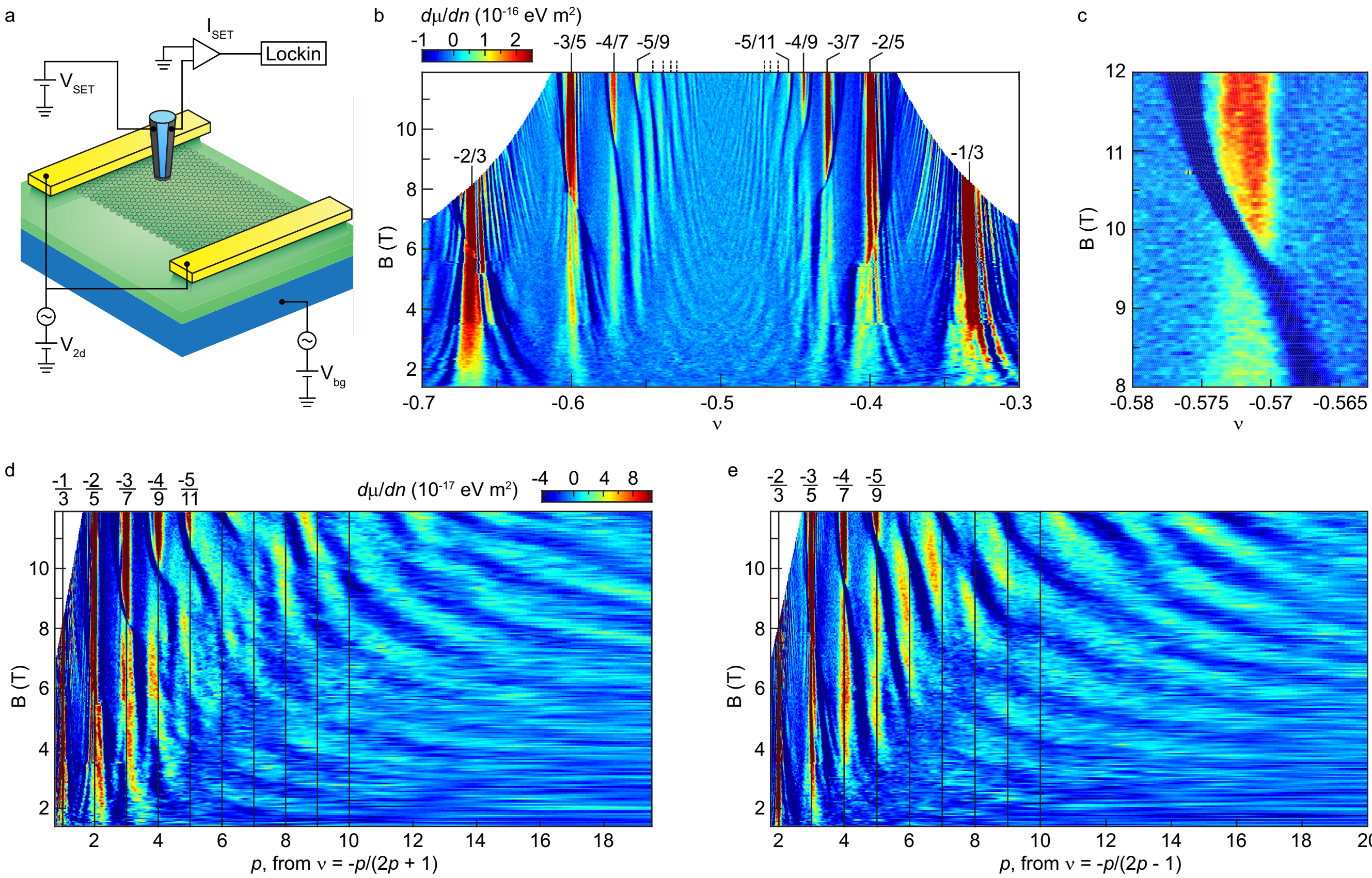

a
$I_{SET}$
Lockin
$V_{SET}$
$V_{2d}$
$V_{bg}$
b
$d\mu/dn$ ($10^{-16}$ eV m$^2$)
-1 0 1 2
-3/5 -4/7 -5/9 -5/11 -4/9 -3/7 -2/5
-2/3
-1/3
B (T)
ν
c
B (T)
ν
d
-1/3 -2/5 -3/7 -4/9 -5/11
$d\mu/dn$ ($10^{-17}$ eV m$^2$)
-4 0 4 8
B (T)
$p$, from $\nu = -p/(2p + 1)$
e
-2/3 -3/5 -4/7 -5/9
B (T)
$p$, from $\nu = -p/(2p - 1)$

Figure 1

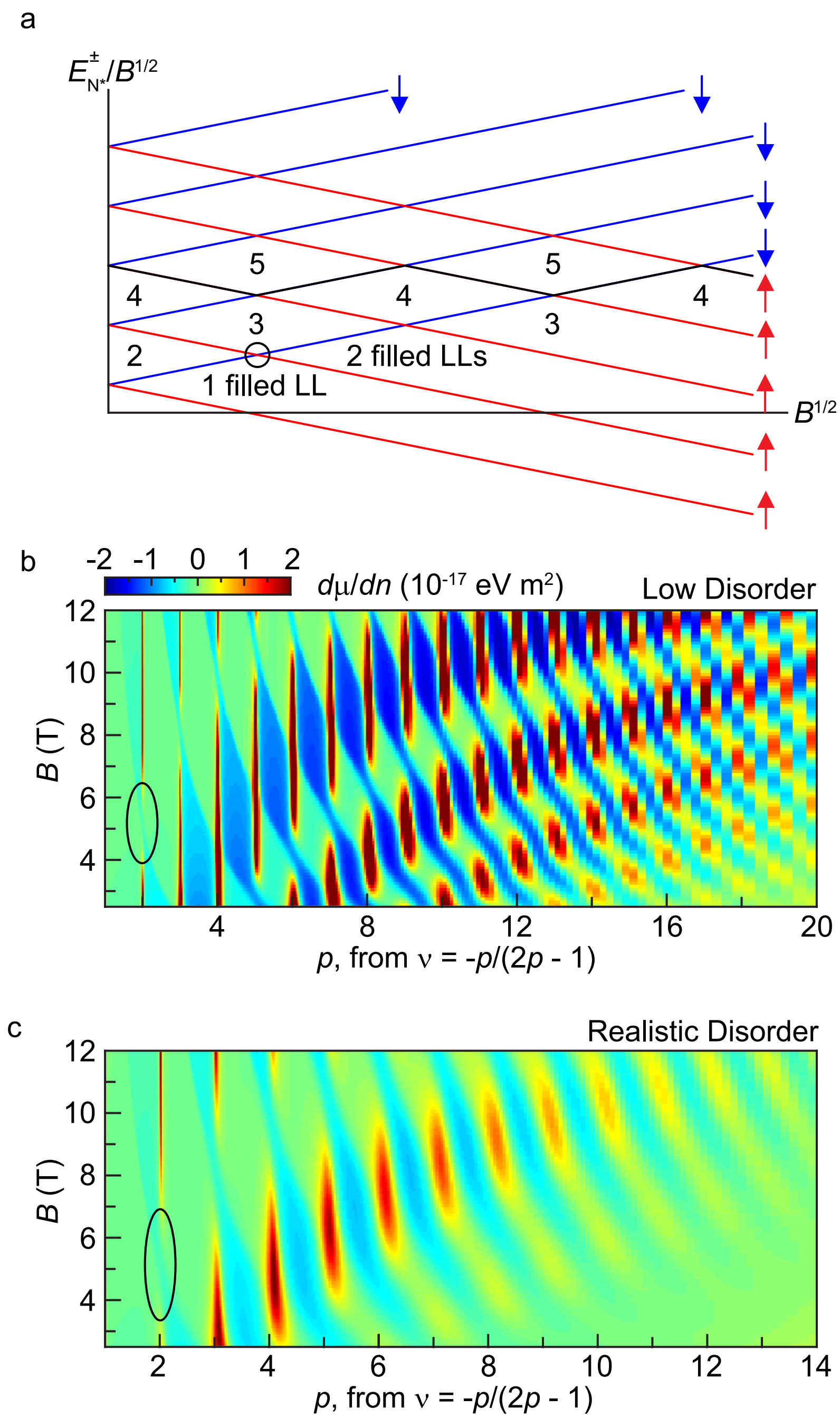

a
E±N*/B1/2
B1/2
5
4
3
2
1 filled LL
2 filled LLs
b
-2 -1 0 1 2
dμ/dn (10-17 eV m2)
Low Disorder
B (T)
p, from ν = -p/(2p - 1)
c
Realistic Disorder
B (T)
p, from ν = -p/(2p - 1)

# Figure 2

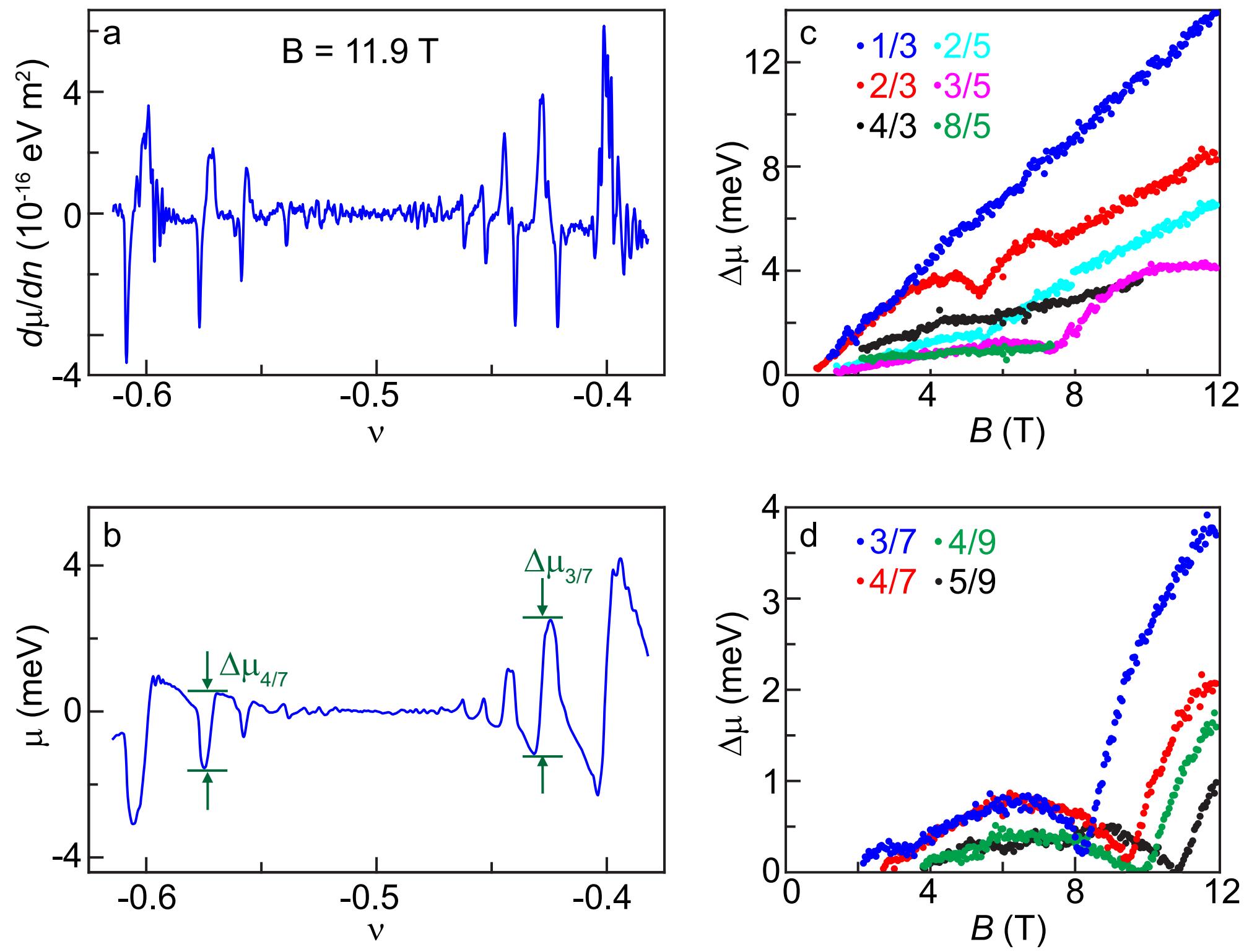

a
B = 11.9 T
dμ/dn (10^-16 eV m^2)
ν
b
μ (meV)
Δμ4/7
Δμ3/7
c
1/3 2/5
2/3 3/5
4/3 8/5
Δμ (meV)
B (T)
d
3/7 4/9
4/7 5/9

Figure 3

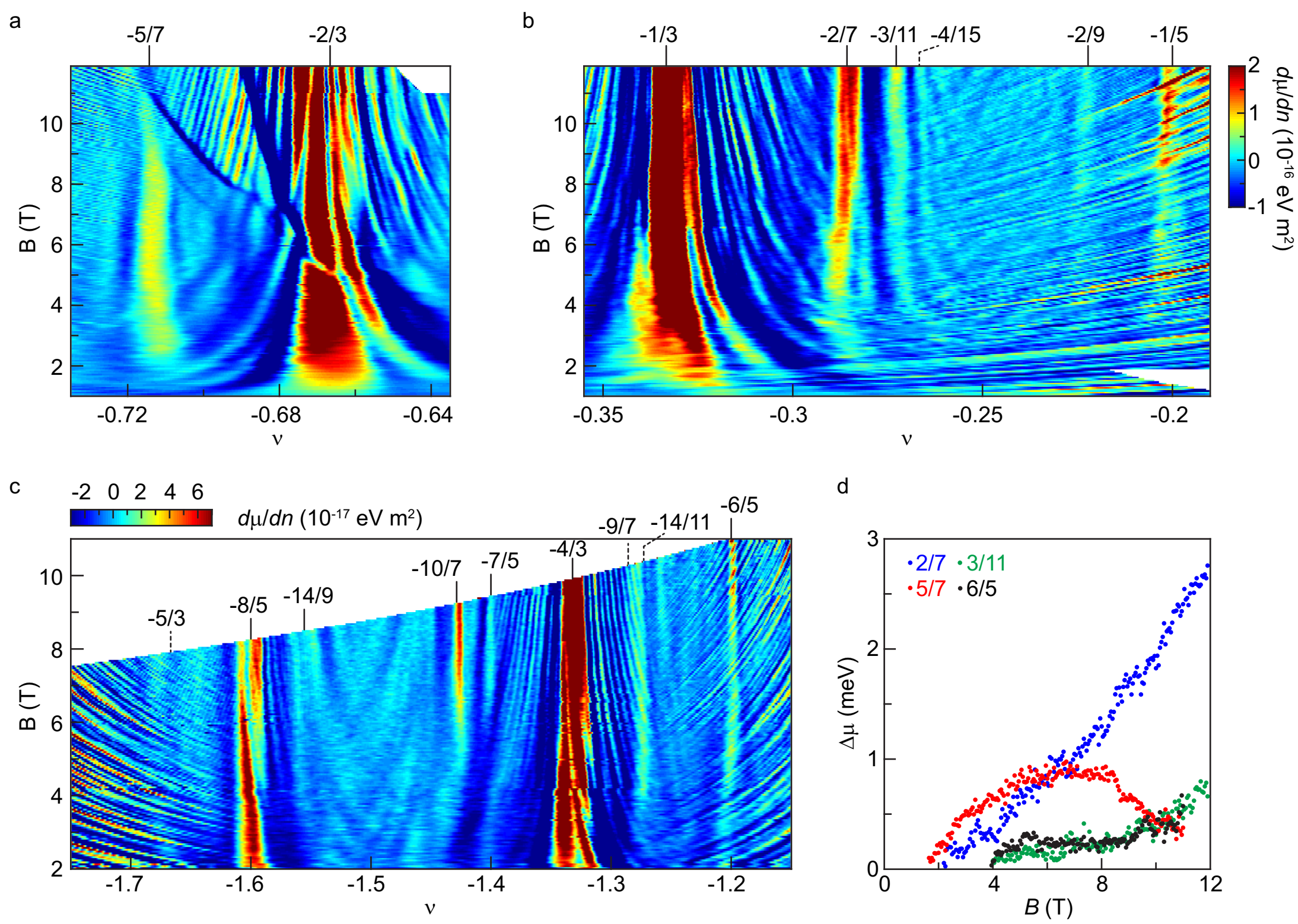

a
-5/7
-2/3
B (T)
ν
-0.72
-0.68
-0.64
b
-1/3
-2/7
-3/11
-4/15
-2/9
-1/5
-0.35
-0.3
-0.25
-0.2
dμ/dn (10^-16 eV m^2)
c
dμ/dn (10^-17 eV m^2)
-5/3
-8/5
-14/9
-10/7
-7/5
-4/3
-9/7
-14/11
-6/5
-1.7
-1.6
-1.5
-1.4
-1.3
-1.2
d
Δμ (meV)
B (T)
2/7
3/11
5/7
6/5

# Figure 4

# Supplementary Information

## Detailed Explanation of Model and Simulation

Below, we describe in detail the method used to numerically simulate the inverse compressibility shown in Fig. 2. The simulation is based on a model of composite fermion Landau level (CF LL) energies $E^{\pm}_{\mathrm{N}^*}$, where $N^*$ is the CF orbital index and ± corresponds to spin-down (up) states. $E^{\pm}_{\mathrm{N}^*}$ contains two terms, the first of which is an effective "kinetic energy" arising from Coulomb interactions

$$E_{\mathrm{N}^*} = \left(N^* - \tfrac{1}{2}\right)\hbar\omega_c^*. \tag{1}$$

In the above expression, $\omega_c^* = {eB^*}/{m^*}$ is the CF cyclotron frequency, $B^* = |B - 2n\Phi_0|$ is the reduced magnetic field felt by CFs, $e$ is the absolute value of the electron charge, $n$ is the carrier density, $B$ is the magnetic field (in Tesla), $\Phi_0 = h/e$ is the flux quantum, and $h = 2\pi\hbar$ is Planck's constant. Because the fractional quantum Hall effect (FQHE) is driven by electron-electron interactions, we assume that the CF effective mass $m^* \propto \sqrt{n}$. Furthermore, we can define filling factor $\nu = nh/eB$, which implies that $B^* = |1 - 2\nu|$ and $m^* \propto \sqrt{\nu B}$ (this expression for mass is similar to that proposed by Park and Jain[1], except for an additional factor of $\sqrt{\nu}$, which is a small factor of order unity and does not dramatically change the qualitative simulated behavior). Substituting these results into Eq. (1) and using the $\nu = p/(2p - 1)$ branch (we use this throughout the derivation because that is what is shown in Fig. 2):

$$E_{\mathrm{N}^*} = \alpha\left(N^* - \tfrac{1}{2}\right)(2\nu - 1)\sqrt{B/\nu}. \tag{2}$$

Both panels in Fig. 2 were generated using fit parameter $\alpha = 1.03\mathrm{x}10^{-22}$ J/T$^{1/2}$, which we determine by matching the phase transitions in our model to the experimental data.

The second term in our model is an energy splitting that scales linearly with magnetic field

$$E^{\pm} = \pm\beta B. \tag{3}$$

In the simplest case, this splitting can arise from the Zeeman effect; then $\beta = 0.5g\mu_B$, where $g \approx 2$ is the $g$-factor of electrons in graphene and $\mu_B$ is the Bohr magneton. For concreteness, we consider this case below. However, our model applies equally well to any form of spin and/or valley energy splitting that scales linearly with magnetic field. The final expression for the energy of CF LLs therefore reads

$$E^{\pm}_{N^*} = E_{N^*} + E^{\pm} = \alpha\left(N^* - \frac{1}{2}\right)(2\nu - 1)\sqrt{\frac{B}{\nu}} \pm \frac{1}{2} g \mu_B B. \tag{4}$$

If we assume that there is no disorder and define $\nu^*$ as the CF filling factor (identical to $p$, except that $\nu^*$ can take on non-integer values), then we can write the following two relations:

$$\frac{d\nu^*}{d\mu} = \sum_{N^*,\pm} \delta(\mu - E^{\pm}_{N^*}) \tag{5}$$

$$\nu^* = \int_{-\infty}^{\mu} \sum_{N^*,\pm} \delta(\mu - E^{\pm}_{N^*})\, d\mu \tag{6}$$

However, in real samples, disorder is always present. We therefore modify Eq. (6) to include finite smearing of the CF LLs:

$$\nu^* = \sum_{N^*,\pm}\left[\frac{1}{\pi}\arctan\left(\frac{\mu - E^{\pm}_{N^*}}{\Delta}\right) + \frac{1}{2}\right] = \frac{\nu}{2\nu - 1}. \tag{7}$$

In the above expression, we have introduced Lorentzian disorder broadening. This represents an arbitrary choice among several possible functional forms that can be used to account for disorder, but using other types of smearing does not dramatically alter the qualitative expectations from the model. In Eq. (7), the parameter Δ can be thought of as the width of the CF LL that arises from disorder, and if we assume a constant density inhomogeneity $\delta n$ in our sample, Δ is a function of $\nu$. We can relate Δ to δ$n$ by taking a derivatives of Eq. (4) as well as the expressions $\nu^* = \nu/(2\nu - 1)$ and $\nu = nh/eB$, yielding:

$$\Delta = \frac{\mathrm{d}E_{\mathrm{N}^*}^{\pm}}{\mathrm{d}N^*}\delta\nu^* = \left[\alpha(2\nu-1)\sqrt{B/\nu}\right]\left[\frac{\delta n\, h}{eB(2\nu-1)^2}\right] = \frac{\delta n(\alpha h)}{e\sqrt{B\nu}(2\nu-1)}. \qquad (8)$$

We use $\delta n = 3\text{x}10^7$ and $1.5\text{x}10^8$ cm$^{-2}$ to produce Figs. 2b and 2c, respectively.

In Eq. (7), the last equality follows from the relation between electron filling factor and the number of occupied CF LLs. Given our expression for $E_{\mathrm{N}^*}^{\pm}$ from Eq. (4), we can numerically solve Eq. (7) for μ as a function of $\nu$, and use simple algebra to recover dμ/d$n$.

**Measurements at Additional Locations**

Although all the data presented in the main text were taken at a single location, the qualitative behavior and phase transitions that we observe are independent of position. Fig. S1 shows inverse compressibility as a function of filling factor and magnetic field at two additional positions that are 300 and 600 nm, respectively, further from the monolayer-bilayer interface than the original measurement position. The phase transitions occur at slightly different critical fields at each location, but the qualitative behavior is unchanged: they are still marked by sharply negative compressibility that cuts across the FQH state, and the critical field increases with filling factor denominator. We emphasize that due to the local nature of the measurement technique, moving our tip to different locations on the flake allows us to effectively probe multiple independent samples.

The variation in the critical fields of the phase transitions suggests that the stability of the FQH states is influenced by the details of the disorder at each location. This is further supported by measurements taken prior to the final current annealing step, when the sample had significantly more disorder. These measurements revealed a $\nu = 2/3$ phase transition around $B$ =

3.5 T at multiple locations (Fig. S2); the large change in critical field likely reflects the large discrepancy in disorder.

**Localized States and Compressibility Oscillations Near $\nu$ = -1/2**

As stated in the main text, localized states are clearly visible near the strongest FQH states (Fig. 1a). Localized states can be simply identified because they run parallel to their parent QH state in the *n*-*B* plane[2], and therefore appear as curved oscillations in compressibility in the $\nu$-*B* plane. Although the modulations in compressibility near $\nu$ = -1/2 also curve in the $\nu$-*B* plane, their origin is distinct. For FQH states with high denominators (*i.e*. between $\nu$ = -4/7 and -5/11), no nearby localized states are visible (Fig. 1c). This is likely because the quasiparticle charge associated with these states is sufficiently small to screen the disorder potential effectively. Due to the absence of localized states surrounding the high-denominator FQH states, and because the oscillations in compressibility near $\nu$ = -1/2 do not match the slope of low-denominator FQH states in the *n*-*B* plane, we conclude that these oscillations are not associated with localized states.

**Zoom-ins on Phase Transitions**

Finer measurements of various phase transitions are depicted in Fig. S3. The pattern of localized states changes above and below several of the phase transitions, indicating that the screening changes, as was observed in GaAs[3]. In some cases, multiple narrow curved negative compressibility features are visible (Figs. S3a, S3b and 4a); this behavior cannot be explained by localized states, and its origin remains unclear. For $\nu$ = -2/5, the phase transition appears to occur virtually instantaneously in magnetic field at the two measurement locations (Figs. S1a and S3c). However, in the third measurement position, the phase transition reverts back to the more typical behavior observed at other filling factors (Fig. S1b). The differences in behavior may be related to the level of disorder at each location; further study is required to fully understand the origin of this behavior.

## Carrier Density at Large Back Gate Voltage

At large back gate voltage $V_{bg}$, the incompressible peaks of the FQH states curve slightly in the $V_{bg}$–$B$ plane. This likely reflects a small change in the capacitance between the back gate and the sample that occurs because the graphene sags slightly at large back gate voltages. To account for this effect, we use the relationship $n = C_g/e[1 + \gamma(V_{bg})^2]V_{bg}$, where $C_g/e = 3.02\text{x}10^{14}$ $m^{-2}$ is proportional to the gate capacitance without any flake sagging and $\gamma = 1.67\text{x}10^{-4}$ is an empirical factor that accounts for the deflection of the sample as the gate voltage is increased (it corresponds to about 3 nm of sagging at $V_{bg} = 10$ V). This correction is used for all figures in the paper, but the impact is only significant for Fig. 4c. The procedure does, however, contribute some uncertainty in the filling factor, particularly at large filling factors ($\nu > 1$) and high magnetic field.

**Figure Legends**

**Figure S1 | Measurements at additional locations. a and b,** Inverse compressibility dμ/d$n$ as a function of filling factor $\nu$ and magnetic field $B$. The measurements were performed at two additional positions 300 nm (a) and 600 nm (b) from those presented in the main text, but the qualitative behavior is similar. Principal FQH states are labeled and marked with black lines.

**Figure S2 | Phase transition at ν = 2/3 before final current annealing step. a-d**, dμ/d$n$ as a function of the carrier density offset from $\nu$ = 2/3 ($\Delta n_{2/3}$) and magnetic field, taken at different locations. The $\nu$ = 2/3 incompressible peak splits into a doublet and weakens considerably between 3 and 3.5 T, but then strengthens again at lower field, indicating a phase transition. The measurement positions in panels (a)-(c) were all separated by at least 300 nm, and the region probed in panel (d) is near to that in (a).

**Figure S3 | Finer measurements of several phase transitions . a-c,** Finer measurements of dμ/d$n$ as a function of $\nu$ and $B$, zoomed in on the phase transitions at $\nu$ = -2/3, -3/5 and -2/5, respectively.

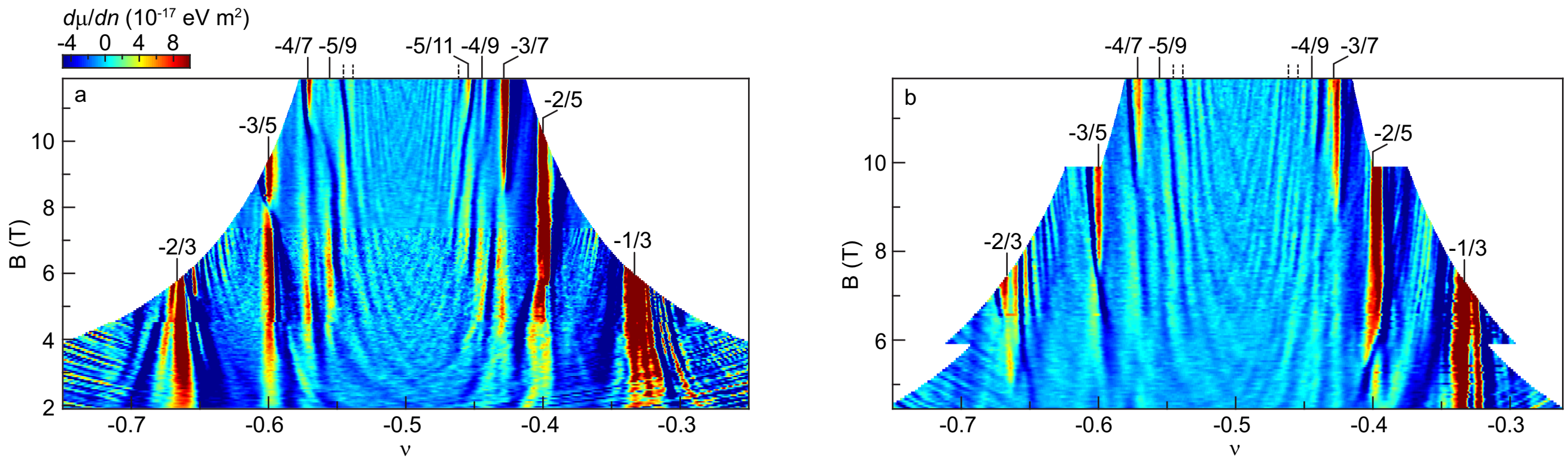

dμ/dn (10^-17 eV m^2)
-4 0 4 8
a
b
-4/7
-5/9
-5/11
-4/9
-3/7
-3/5
-2/5
-2/3
-1/3
B (T)
ν
-0.7
-0.6
-0.5
-0.4
-0.3

# Figure S1

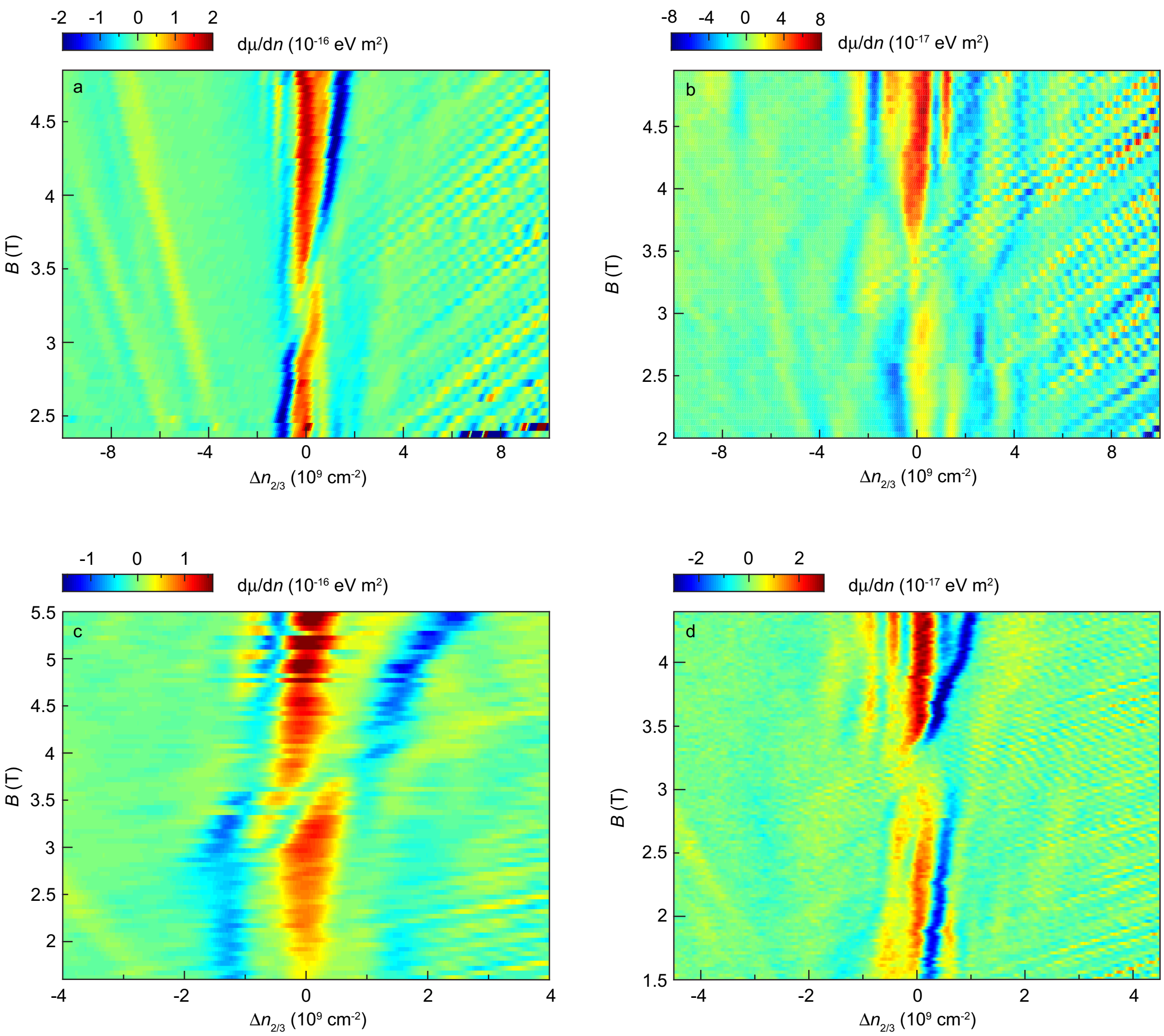

-2 -1 0 1 2
dμ/dn (10^-16 eV m^2)
a
B (T)
Δn_2/3 (10^9 cm^-2)
-8 -4 0 4 8
dμ/dn (10^-17 eV m^2)
b
B (T)
Δn_2/3 (10^9 cm^-2)
-1 0 1
dμ/dn (10^-16 eV m^2)
c
B (T)
Δn_2/3 (10^9 cm^-2)
-2 0 2
dμ/dn (10^-17 eV m^2)
d
B (T)
Δn_2/3 (10^9 cm^-2)

# Figure S2

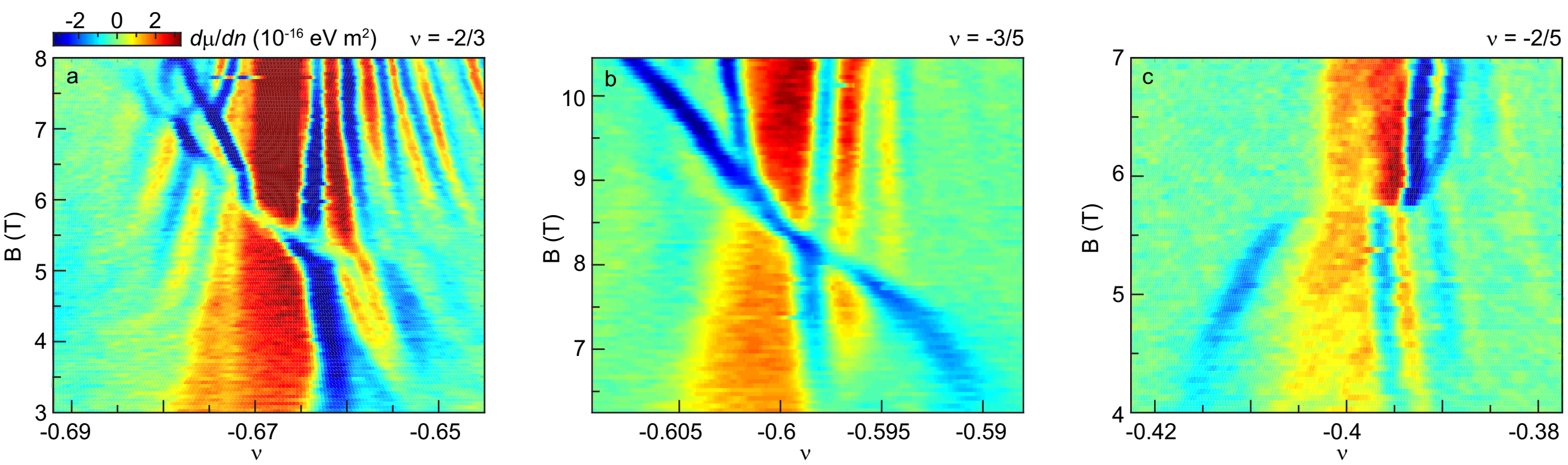

# Figure S3